# Narrowband telecom band polarization-entangled photon source by superposed monolithic cavities

Ming-Yuan Gao[1,2,4], Yin-Hai Li[1, 2, 4, *], Yan Li[1, 2, 3], Zhenghe Zhou[1], Guang-Can Guo[1, 2, 3], Zhi-Yuan Zhou[1, 2, 3,] * and Bao-Sen Shi[1, 2, 3, 5]

[1] CAS Key Laboratory of Quantum Information, University of Science and Technology of China, Hefei, Anhui 230026, China
[2] CAS Center for Excellence in Quantum Information and Quantum Physics, University of Science and Technology of China, Hefei 230026, China
[3] Hefei National Laboratory, University of Science and Technology of China, Hefei 230088, China
[4] These two authors contributed equally to this article.
[5] drshi@ustc.edu.cn
*Corresponding authors: lyhly@ustc.edu.cn; zyzhouphy@ustc.edu.cn

A high-quality narrowband polarization-entangled source in the telecom band is preferred to avoid frequency dispersion for long-distance transmission in optical fibers and to efficiently couple with telecom band quantum memories. Here, we report narrowband, telecom-band, polarization-entangled photon pair generation based on the superposition of single-longitudinal-mode photon pairs from two monolithic nonlinear crystal cavities in a passively stable interferometer based on beam displacers. The photon pairs generated from the cavities exhibit a high coincidence to accidental coincidence ratio of 20000 and a bandwidth below 500 MHz. Two-photon polarization interference, Bell-inequality, and quantum state tomography are performed to indicate the high quality of the entangled source. The current configuration demonstrates greater stability than traditional free space cavity-enhanced polarization-entangled state generation, which is promising for quantum communication applications.

For long-distance entanglement distribution, the temporal duration of the photon will be broadened due to frequency dispersion in optical fibers [1, 2]. A narrowband polarization-entangled photon source is preferred in this application. To construct a global quantum communication network [3], quantum memories are indispensable components for a quantum repeater [4]. However, most quantum memories require photonic states with bandwidths below GHz for effective light-matter interactions [5-8]. Spontaneous parametric down-conversion (SPDC) in single-pass bulk crystals usually has a bandwidth of more than 100 GHz, which is far beyond the requirement of some quantum communication applications [9]. To address this issue, cavity-enhanced SPDC has been proposed for decades [10]. In cavity-SPDC, the bandwidth of the photon pairs is determined by the bandwidth of the cavity. Substantial progress has been made since the initial demonstration [11-18]. For instance, ultra-high brightness, narrowband photon pairs have been prepared in triple-resonant configurations [11,12]. A two-color heralded single photon source with one photon in the telecom band and the other in the memory band and a polarization-entangled photon source have been prepared [13,14]. Single-longitudinal-mode photon pairs can be generated directly from a monolithic cavity based on frequency cluster effects in type-II phase matching [15-18].

Previous demonstrations have focused primarily on preparing narrowband heralded single-photon sources [11-13, 15-18], and only a few demonstrations have been conducted on a narrowband polarization-entangled photon source [14]. Moreover, existing demonstrations of single-longitudinal-mode, narrowband, polarization-entangled photon sources based on free space cavities are very complicated, and require complex auxiliary optical and electrical control systems to lock the cavity [14]. Based on the intrinsic single-longitudinal-mode emission of photon pairs generated in two monolithic cavities formed by two type-II periodically poled potassium titanyl phosphate (PPKTP) crystals, a high-quality, single-longitudinal-mode, narrowband, polarization-entangled photon source obtained by superposing two-photon pairs in a passively stable interferometer formed by beam displacers (BDs) is reported. The spectral bandwidth of the source is less than 500 MHz, and the maximal coincidence to accidental coincidence ratio (CAR) is greater than 20000. The achieved entanglement qualities include: two-photon interference visibilities in 0º and 45º bases are $(99.97 \pm 0.03)$ % and $(87.09 \pm 0.72)$ %, respectively; the obtained Clauser–Horne–Shimony–Holt (CHSH) Bell S parameter is $S=2.639 \pm 0.048$; and the fidelity of the experimentally reconstructed density matrix is $0.907 \pm 0.006$. All of these values indicate that the source is of high quality and will have broad applications in quantum communications.

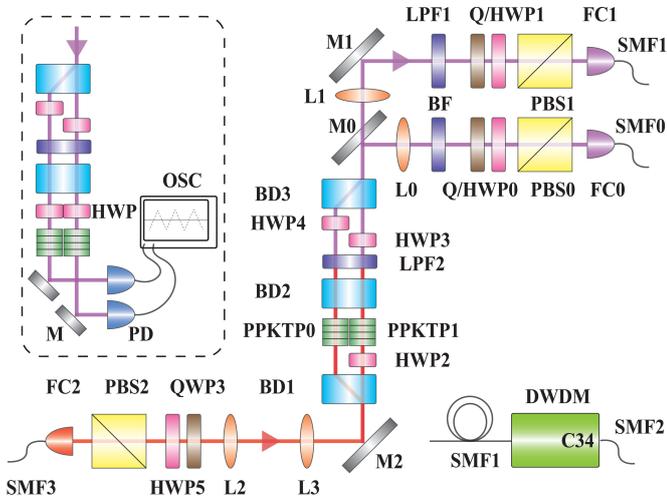

Fig. 1. Experimental setup. Inset shows the setup for measuring the spectral linewidth for each of the polarizations modes (see Fig. 2). SMF: single-mode fiber; FC: fiber coupler; PBS: polarization beam splitter; Q(H)WP: quarter-(half-) wave plate; L: lens; M: mirror; BD: beam displacer; PPKTP: periodically poled KTP crystal; LPF: long-pass filter; BF: bandpass filter; DWDM: dense wave division multiplexing; PD: photodiode; OSC: oscilloscope;

The experimental setup is shown in Fig. 1. Two PPKTP crystals with nearly the same specifications are used to generate type-II degenerate SPDC photons. The crystal has dimensions of 1 mm (thickness) × 2 mm (width) × 1.47 mm (length) and a poling period of 46.2 μm. The beam propagated along the x-axis of the crystal. The horizontal (H) and vertical (V) polarizations correspond to the y and z axes of the crystal, respectively. The crystal has a high-reflection coating at 1550 nm and an antireflection coating at 775 nm on one face, while the other face has a partially reflective coating at 1550 nm (the reflectivity equals 96%) and an antireflection coating at 775 nm. The two surfaces of the crystal therefore form a Fabry-Pérot (FP) cavity where photons resonate [18]. A continuous-wave laser at 775.1 nm (TA pro, TOPTICA, linewidth <1 MHz) was used as a pump beam, which was output from FC2, and was incident on BD1 (further details about BD are provided in the Appendix) through a lens group (150 mm - 50 mm). BD1 separated the input pump beam into H and V output pump beams. The V pump beam was converted into the H pump beam after passing through a half-wave plate. The two beams passed through two separate temperature-controlled crystals (temperature stability of $\pm 1$ mK, Anhui KunTeng Quantum Technology, Co. Ltd.) to produce type-II cavity-enhanced SPDC. The photon polarization states generated by PPKTP0 (PPKTP1) are labeled H0 and V0 (H1 and V1). The H and V photons were separated perpendicularly to the top and bottom of the paper after passing through BD2. Thus far, four photons had propagated separately. The photon polarization state V0 (H1) in the upper left (lower right) was adjusted to H0 (V1) by the half-wave plate. The polarization states of the other two photons H0 and V1 remained unchanged. Subsequently, the H and V photons at the same height converged after four photons passed through BD3. The process of photon transport from BD1 to BD3 is shown in Fig. A1 in the Appendix. After BD3, the photons in the upper part passed over M0 and were finally collected into FC1, while the photons in the lower part were reflected by M0 and finally collected into FC0. SMF1 was connected to the C34 channel (with a center wavelength of 1550.12 nm) of a 200 GHz spacing dense wave division multiplexing to retain the single-longitudinal mode and remove a small amount of other low-intensity longitudinal modes. SMF0 and SMF2 were connected to the superconductor nanowire single-photon detectors, and finally, the coincidence measurement was performed with a time window of 3.2 ns and a cumulative time of 10 s.

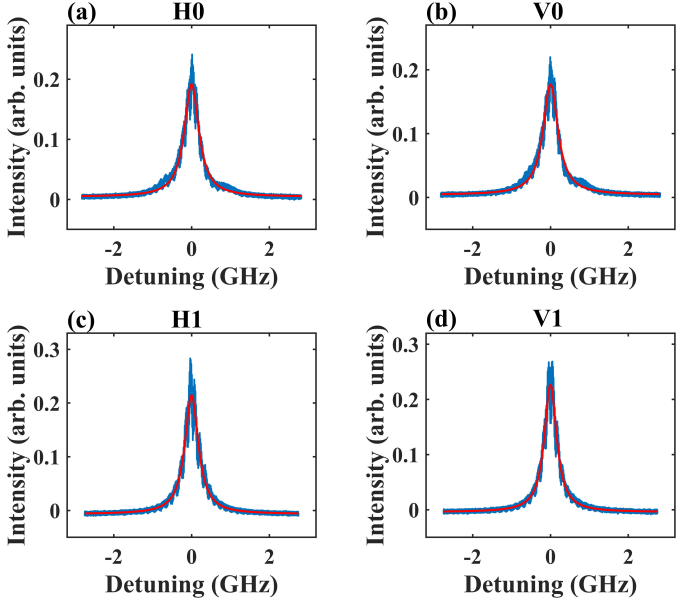

Fig. 2. Bandwidths of the (a) and (c) horizontal polarization modes and (b) and (d) vertical polarization modes of PPKTP0 and PPKTP1, respectively. The blue parts are the raw data from the oscilloscope. The fringes might be caused by cavity temperature instability. The red curves are the results of fitting these data using the Lorentzian function.

First, a wavelength-tunable infrared laser (CTL 1550, TOPTICA) was used to characterize the two FP cavities. The laser was output from FC0 and reflected by M0, and the remaining setup is shown in the inset of Fig. 1 (the unlabeled components are the same as those shown outside the inset). The transmission spectrum was obtained by scanning the piezoelectric transducer (PZT) voltage of the CTL 1550 (the PZT voltage linearly corresponds to the laser wavelength) using a photodiode and an oscilloscope, as shown in Fig. 2. The angle of the half-wave plate in front of the PPKTP is adjusted to switch the cavity spectrum in V or H polarization mode. The H polarization mode bandwidth of PPKTP0 (PPKTP1) is 454 MHz (422 MHz), and the V polarization mode bandwidth is 462 MHz (384 MHz). The free spectral range (FSR) of the H polarization mode is 57.91 GHz (57.41 GHz) and the FSR of the V polarization mode is 54.91 GHz (54.91 GHz). The FSR difference between the two modes is 3 GHz (2.5 GHz), which is greater than the average bandwidth of 0.459 GHz (0.403 GHz). This ensures that there will be only a single mode in a cluster [15].

The crystal cluster spacing is 1.06 THz (1.26 THz), which is greater than half of its bandwidth (the full width at half maximum is 2.04 THz).

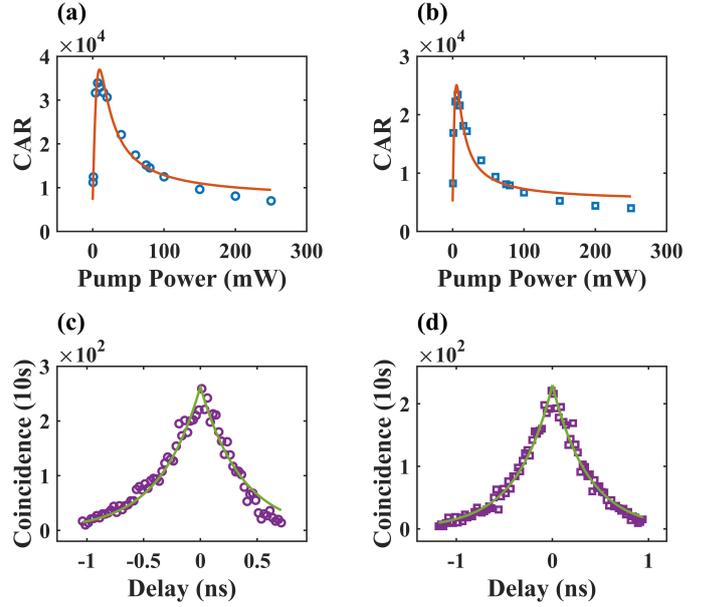

Fig. 3. (a) and (b) are coincidence to accidental coincidence ratios (CARs) corresponding to PPKTP0 and PPKTP1, respectively; (c) and (d) are second-order cross-correlation functions $g^{(2)}$ corresponding to PPKTP0 and PPKTP1, respectively.

Next, we characterize the quantum properties of each crystal. By changing the pump power, coincidence counts and accidental counts were obtained at the different pump powers, and the CAR was obtained accordingly, as shown in Fig. 3(a) and (b). The fitting model provided in Ref. [19] is used. At a low pump power, the photon generation rate increases and the effective coincidence counts relative to the dark counts increase, leading to an increase in the CAR. However, at a high generation rate, the obvious multiphoton effect results in a decrease in the CAR. The maximal CAR of PPKTP0 (PPKTP1) exceeds $3\times10^4$ ($2\times10^4$), and the CAR exceeds $6\times10^3$ ($4\times10^3$) in the power range of 0.5 mW-250 mW. We then measured the second-order cross-correlation function $g^{(2)}$ of each crystal using a pump power of 75 mW and a coincidence resolution of 25 ps. The results are shown in Fig. 3(c) and (d). The fitting function is $e^{-2\pi\gamma|\tau|}$. Based on the fitting curve, the $g^{(2)}$ of PPKTP0 (PPKTP1) has a full width at half maximum (FWHM) value of $0.502 \pm 0.022$ ns ($0.519 \pm 0.025$ ns). The time jitter of the detectors and the coincidence system is approximately 60 ps. The FWHM of $g^{(2)}$ is defined as $T_{FWHM} = 1.39/2\pi\gamma'$, where $\gamma'$ is the geometric mean of $\gamma_s$ and $\gamma_i$. In our cavity, $\gamma_s$ and $\gamma_i$ are 454 MHz and 462 MHz (422 MHz and 384 MHz) respectively; thus, the calculated FWHM is 0.483 ns (0.550 ns), which agrees with the value obtained from $g^{(2)}$. The above FWHM data indicate that single-longitudinal-mode narrowband photon pairs are generated [18].

We then characterize the polarization entanglement. After the photons pass through BD3, the polarization-entangled state is $|\Phi\rangle = (|HH\rangle + e^{i\theta}|VV\rangle)/\sqrt{2}$, where $\theta$ can be modified by adjusting the phase of the pump (by HWP5 and QWP3) or fine-tuning the angle of the BDs. Finally, the state $|\Phi^-\rangle = (|HH\rangle - |VV\rangle)/\sqrt{2}$ is generated by changing the relative phase $\theta$ of the state $|\Phi\rangle$. The pump power was set to 150 mW and a polarization correlation measurement was first performed. The visibilities in 0° and 45° bases are (99.97 ± 0.03) % and (87.09 ± 0.72) % respectively, as shown in Fig. 4(a). The reason for the decrease in visibility at 45° basis is mainly because of the incomplete overlap of the spectra and temporal of the photons generated by the two crystals. The spectral and temporal overlap can be estimated by $R_{overlap} = T_1/T_0 = \sqrt{\gamma_{s0}\gamma_{i0}}/\sqrt{\gamma_{s1}\gamma_{i1}} \approx 0.88$, where $T_0$ ($T_1$) is the FWHM of the $g^{(2)}$ of PPKTP0 (PPKTP1). The maximal visibility at 45° basis is equal to the spectral and temporal overlap. The visibility at 45° basis can be improved by using an FP cavity with a bandwidth of less than 384 MHz. The estimated total photon collection efficiency (including transmittance loss, coupling efficiency, and detection efficiency) is 12.5%, so the emission spectral brightness is 0.7 (s· mW· MHz)$^{-1}$. Next, we performed the CHSH–Bell inequality test [20] and the result obtained is S=2.639 ± 0.048, which violates the Bell inequality by 13 standard deviations. Finally, we performed quantum state tomography. The reconstructed real and imaginary parts of the density matrix are shown in Fig. 4(b) and (c), respectively, and the fidelity compared with the ideal state is 0.907 ± 0.006 [21]. It is worth mentioning that all the error estimates are obtained by assuming a Poisson distribution of the data.

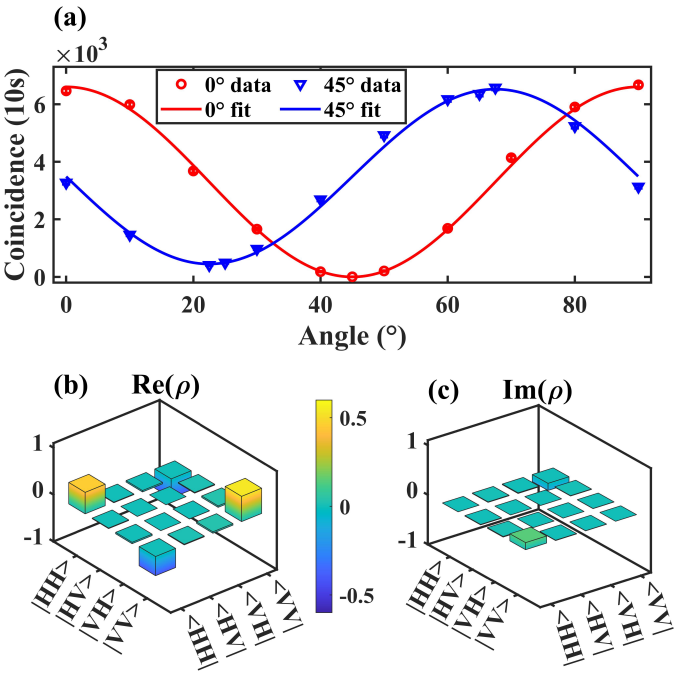

Fig. 4. Polarization entanglement characterization for the produced state. (a) Polarization interference curves in 0° and 45° bases; (b) and (c) are the reconstructed real and imaginary parts of the density matrix $\rho$, respectively.

In summary, a high-quality, single-longitudinal-mode, narrowband, polarization-entangled photon source is prepared here by combining a passively stable interferometer (the interferometer was placed on our laboratory optical table and no other steps were taken to protect it from environmental perturbations.) and SPDC in two monolithic cavity-enhanced SPDC using two type-II PPKTP crystals. The source has a bandwidth of less than 500 MHz and high CARs. The entanglement quality was characterized by various methods, and the quality of entanglement can be further improved with narrowband FP filters to increase the spectral and temporal overlap between the photon pairs generated in different monolithic cavities. The present work offers an effective method for generating high-quality, narrowband, polarization-entangled photon sources, which is very promising for quantum communication applications.

This work is supported by the National Natural

Appendix: Details about the beam displacers

The beam displacer (BD) can separate an input beam into two orthogonal linearly polarized output beams or combine two orthogonal linearly polarized input beams. The beam displacement is approximately 4 mm. The beam displacement in the horizontal or vertical plane can be achieved by rotating the optical axis of the BD. BD1 (BD2 and BD3) is (are) coated with an antireflection coating at 775 nm (1550 nm). The process of photon transport from BD1 to BD3 is shown in Fig. A1 (The long-pass filter is not shown). The blue arrows on the BDs indicate the direction of beams that separate or converge within them.

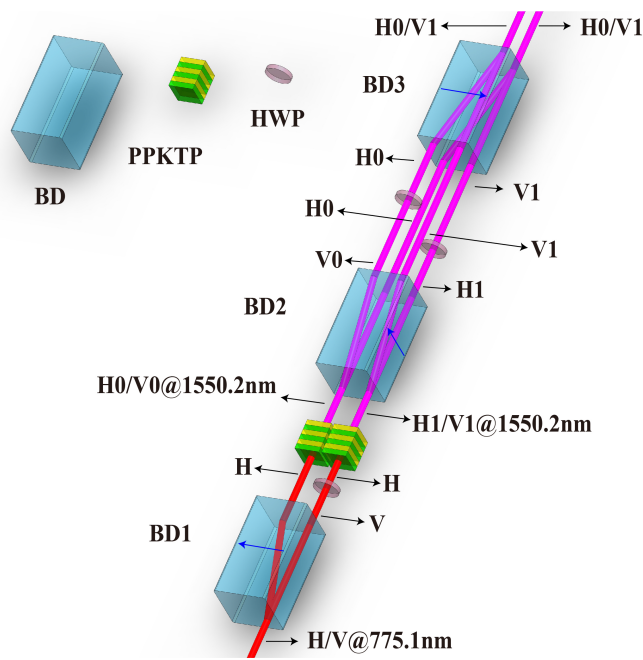

Fig. A1. Experimental setup of the beam displacers. BD: beam displacer; PPKTP: periodically poled KTP crystal; HWP: half-wave plate;

1. Q. Zhang, H. Takesue, S. W. Nam, C. Langrock, X. Xie, B. Baek, M. M. Fejer, and Y. Yamamoto, Distribution of Time-Energy Entanglement over 100 km fiber using superconducting single-photon detectors, Opt. Express 16, 5776(2008).
2. J. F. Dynes, H. Takesue, Z. L. Yuan, A. W. Sharpe, K. Harada, T. Honjo, H. Kamada, O. Tadanaga, Y. Nishida, M. Asobe, and A. J. Shields, Efficient entanglement distribution over 200 kilometers, Opt. Express 17,11440(2009).
3. H. J. Kimble, The quantum internet, nature, 453, 1023(2008).
4. B.-S. Shi, D.-S. Ding and W. Zhang, Quantum storage of orbital angular momentum entanglement in cold atomic ensembles, J. Phys. B: At. Mol. Opt. Phys. 51, 032004 (2018).
5. A. I. Lvovsky, B. C. Sanders and W. Tittel, Optical quantum memory, Nat. Photon. 3, 706(2009).
6. E. Saglamyurek, J. Jin, V. B. Verma, M. D. Shaw, F. Marsili, S. W. Nam, D. Oblak and W. Tittel, Quantum storage of entangled telecom-wavelength photons in an erbium-doped optical fibre, Nat. Photon. 9, 83(2015).
7. J. Jin,1, E. Saglamyurek, M.. lí G. Puigibert, V. Verma, F. Marsili, S. W. Nam, D. Oblak, and W. Tittel, Telecom-Wavelength Atomic Quantum Memory in Optical Fiber for Heralded Polarization Qubits, Phys. Rev. Lett. 115, 140501 (2015).
8. K. Shinbrough, D. R. Pearson, B. Fang, E. A. Goldschmidt, and V. O. Lorenz, Broadband quantum memory in atomic ensembles, Advances in Atomic, Molecular and Optical Physics 72: 297-360 (2023).
9. S. Wengerowsky, S. K. Joshi, F. Steinlechner, H. Hübel, and R. Ursin, An entanglement-based wavelength-multiplexed quantum communication network, Nature 564, 225 (2018).
10. Z. Y. Ou and Y. J. Lu, Cavity Enhanced Spontaneous Parametric Down-Conversion for the Prolongation of Correlation Time between Conjugate Photons, Phys. Rev. Lett. 83, 2556-2559 (1999).
11. Z.-Y. Zhou, D.-S. Ding, Y. Li, F.-Y. Wang, and B.-S. Shi, Cavity-enhanced bright photon pairs at telecom wavelengths with a triple-resonance configuration, J. Opt. Soc. Am. B 31, 128-134 (2014).
12. M. Scholz, L. Koch, and O. Benson, Statistics of Narrow-Band Single Photons for Quantum Memories Generated by Ultrabright Cavity-Enhanced Parametric Down-Conversion, Phys. Rev. Lett. 102, 063603 (2009).
13. J. Fekete, D. Rieländer, M. Cristiani, and H. de Riedmatten, Ultranarrow-Band Photon-Pair Source Compatible with Solid State Quantum Memories and Telecommunication Networks, Phys. Rev. Lett. 110, 220502 (2013).
14. J. Wang, Y.-F. Huang, C. Zhang, J.-M.Cui, Z.-Y. Zhou, B.-H. Liu, Z.-Q. Zhou, J.-S. Tang, C.-F. Li, and G.-C. Guo, Universal Photonic Quantum Interface for a Quantum Network, Phys. Rev. Appl. 10, 054036 (2018).
15. E. Pomarico, B. Sanguinetti, C. I. Osorio, H. Herrmann, and R. T. Thew, Engineering integrated pure narrow-band photon sources, New J. of Phys. 14, 033008(2012).
16. P. Enrico, S. Bruno, G. Nicolas, T. Robert, Z. Hugo, S. Gerhard, T. Abu, and S. Wolfgang, Waveguide-based OPO source of entangled photon pairs, New J. of Phys. 11, 113042 (2009).
17. C.-S. Chuu, G. Y. Yin, and S. E. Harris, A miniature ultrabright source of temporally long, narrowband biphotons, Appl. Phys. Lett. 101, 051108 (2012).
18. Y.-H. Li, Z.-Y. Zhou, S.-L. Liu, Yan Li, S.-K. Liu, C. Yang, S. Wang, Z.-H. Zhu, W. Gao, G.-C. Guo, and B.-S. Shi, Compact sub-GHz bandwidth single-mode time-energy entangled photon source for high-speed quantum networks, OSA Continuum 4, 608(2021).
19. K. Harada, H. Takesue, H. Fukuda, T. Tsuchizawa, T. Watanabe, K. Yamada, Y. Tokura, and S. Itabashi, Frequency and Polarization Characteristics of Correlated Photon-Pair Generation Using a Silicon Wire Waveguide, IEEE J. Sel.Top. Quantum Electron. 16(1), 325–331 (2010).
20. J. F. Clauser, M. A. Horne, A. Shimony, and R. A. Holt, Proposed experiment to test local hidden-variable theories, Phys. Rev. Lett.23, 880–884 (1969).
21. D. F. V. James, P. G. Kwiat, W. J. Munro, and A. G. White, Measurement of qubits, Phys. Rev. A 64(5),052312 (2001).